\documentclass[11pt,twoside]{article}

\usepackage{asp2006}
\usepackage{epsf}
\usepackage{psfig}
\usepackage{lscape}

\begin{document}
\title{The Physical Properties of Red Supergiants}   
\author{Philip Massey,\altaffilmark{1}
Bertrand Plez,\altaffilmark{2} 
Emily M. Levesque,\altaffilmark{3}
K. A. G. Olsen,\altaffilmark{4}
David R. Silva,\altaffilmark{5}
and Geoffrey C. Clayton\altaffilmark{6}}
\altaffiltext{1}{Lowell Observatory, 1400 W. Mars Hill Rd., Flagstaff, AZ 86001}
\altaffiltext{2}{GRAAL, Universit\'{e} de Montpellier II, SNRC, 34095 Montpellier, France}
\altaffiltext{3}{Institute for Astronomy, University of Hawaii, 2680 Woodlawn Drive, Honolulu, HI 96822}
\altaffiltext{4}{Gemini Science Center, NOAO, P. O. 26732, Tucson, AZ 85726-6732}
\altaffiltext{5}{Thirty Meter Telescope, 2636 E. Washington Blvd., Pasadena, CA 91107}
\altaffiltext{6}{Department of Physics \& Astronomy, Louisiana State University, Baton Rouge, LA 70803}

\begin{abstract} 
Red supergiants (RSGs)
are an evolved stage in the life of intermediate massive stars ($<25M_\odot$).
For many years their location in the H-R diagram was at variance with the evolutionary models.
Using the MARCS stellar atmospheres, we have determined new effective temperatures and
bolometric luminosities for RSGs
in the Milky Way, LMC, and SMC, and our work has resulted in much better agreement with
the evolutionary models.  We have also found evidence of significant visual extinction 
due to circumstellar dust.  Although in the Milky Way the RSGs contribute only a small fraction
($<$1\%) of the dust to the interstellar medium (ISM), in starburst galaxies or galaxies at large
look-back times, we expect that RSGs may be the main dust source.  We are in the process of
extending this work now to RSGs of higher and lower metallicities using the galaxies M31 and WLM.

\end{abstract}


\section{Introduction: What Am I Doing Here?}   

\begin{figure}[!ht]
\plotfiddle{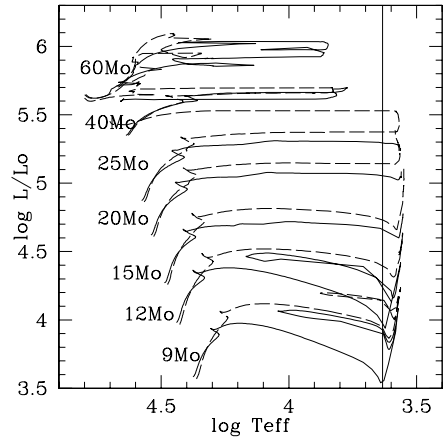}{2.0in}{0.}{30}{30}{-80}{-50}
\caption{\label{fig:hrd} The evolutionary tracks of Meynet \& Maeder (2003),
corresponding to solar metallicity,  are shown.  Solid curves are for stars with
no initial rotation, while dashed curves are for stars with an initial rotation velocity
of 300 km s$^{-1}$.  The solid vertical line corresponds to an effective temperature of 4300 K,
cooler than which a star is spectroscopically identified as a K-M.
}
\end{figure}

I feel I have to explain what I'm doing at a ``cool stars" 
conference, much less giving the opening talk.  Most of my work has been
concerned with the formation, physical
properties, and evolution of massive stars, and for most of my career I've stayed 
over on
the ``hot" side of the H-R diagram (HRD).  Still, as shown in Figure~\ref{fig:hrd} {\it some} massive
stars spend a significant fraction of their lives as red supergiants (RSGs).  For these particular
tracks (computed for Galactic metallicity and including the effects of rotation)  stars of $20 M_\odot$ and lower spend the majority of their He-burning lives as RSGs.
Stars of 25$M_\odot$ spend some of their He-burning phase as RSGs, but then
may evolve back to the blue side of the HRD, where they end
their lives as Wolf-Rayet stars (WRs).  Higher mass stars never make it over to the 
RSG side.

The fraction of the He-burning life spent as a RSG, as well as the mass limits for
becoming RSGs, should depend upon metallicity: in lower metallicity environments,
such as the SMC, we expect to find lots of RSGs and few WRs, while in higher
metallicity environments the opposite should be true.  This was first pointed out by
Maeder, Lequeux, \& Azzopardi (1980).  So, the relative number of RSGs and WRs provide a sensitive test of stellar evolutionary theory.

\begin{figure}[!ht]
\plotfiddle{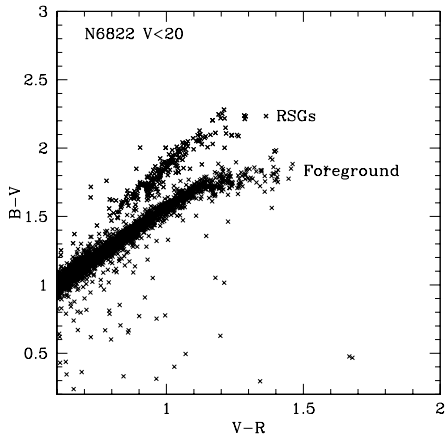}{2.0in}{0.}{30}{30}{-80}{-50}
\caption{\label{fig:2color} This two-color plot allows us to distinguish foreground dwarfs
from bona fide RSGs for the nearby galaxy NGC 6822.  The data are from Massey et al.\ (2007b).
}
\end{figure}

Still, first you have to find them.  My interest in the subject of RSGs was first piqued by the comparison of blue and red stars in the nearby galaxy M33 given by Humphreys \& Sandage (1980).  They were nowhere near alike!  The blue stars were mostly clumped
together in what these authors designated OB associations, while the red stars showed
a fairly uniform distribution across the face of the galaxy. (Compare their Figures 21 and 22.)  They correctly interpreted this as the result of contamination by foreground Galactic K and M dwarfs.  A little fiddling with the best (at the time) model atmospheres
convinced me that you {\it could} separate out the two populations by
a $B-V$, $V-R$ two color diagram (Massey 1998).  Figure~\ref{fig:2color} shows such a diagram for stars seen towards NGC~6822, a Local Group irregular galaxy at fairly low Galactic latitude ($b=-18.4^\circ$).

\begin{figure}[!ht]
\plotfiddle{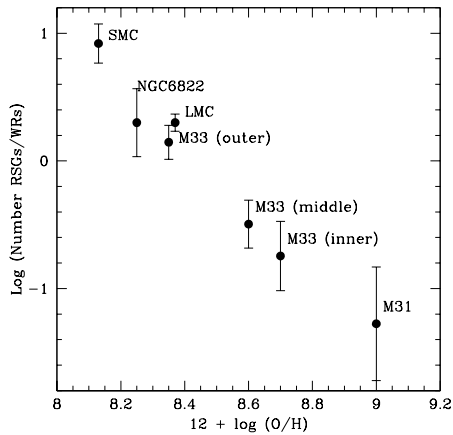}{2.0in}{0.}{30}{30}{-80}{-50}
\caption{\label{fig:RSGWRs} The ratio of the number of RSGs to WRs correlates nicely
with metallicity within the local group.  This figure is based upon one in Massey (2003a).
}
\end{figure}

We eventually carried out RSG searches in selected regions of NGC 6822, M33, and M31 (Massey 1998) and the SMC and LMC (Massey \& Olsen 2003).  At the time we
came up with an amazingly nice correlation of the RSG/WR number ratio with metallicity, which goes in the same sense as expected (Figure~\ref{fig:RSGWRs}).  The number ratio changes by a factor of 100 or so over just 0.9~dex in the log of the oxygen abundances.

\begin{figure}[!ht]
\plottwo{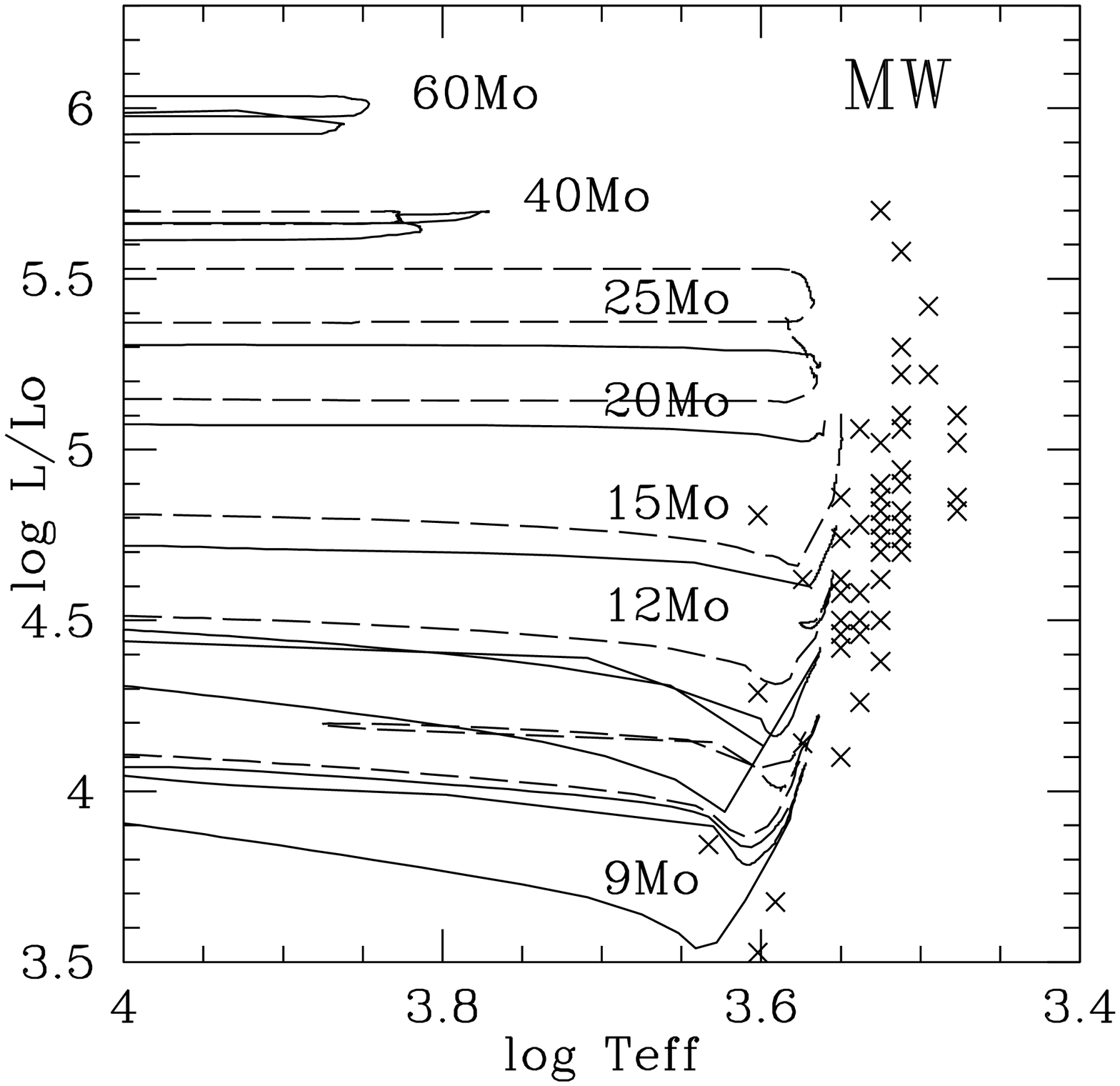}{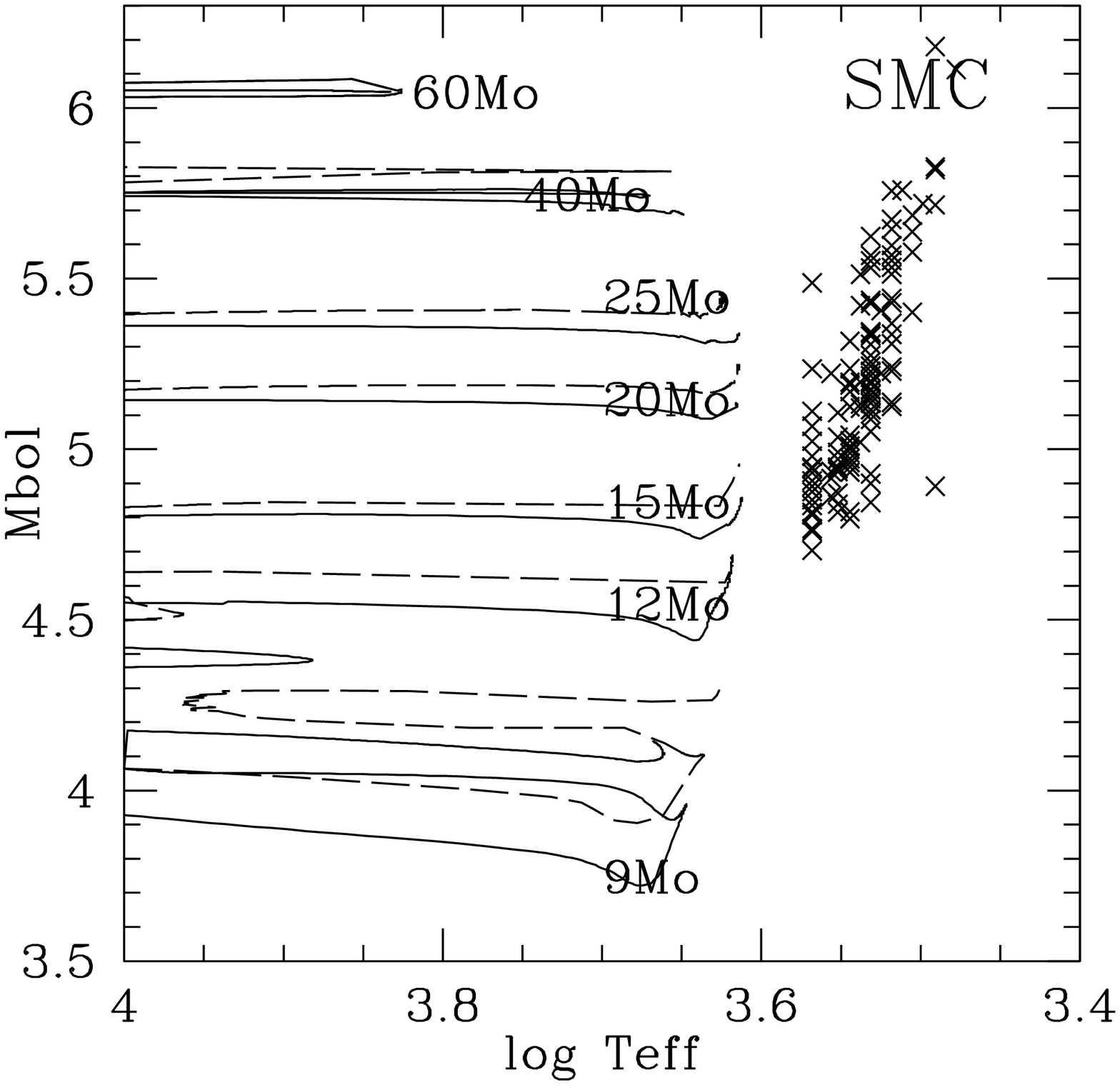}
\caption{\label{fig:HROLD}  Massey (2003a) and Massey \& Olsen (2003)
found that red supergiants were considerably cooler and more luminous than the evolutionary tracks predicted.  On the left we show the comparison for Milky Way
RSGs, where the 
data are from Humphreys (1978), and the evolutionary tracks are those of
Meynet \& Maeder (2003) for a Galactic metallicity ($z=0.020$).  On the
right we show a similar comparison for the SMC, where the data come
from Massey \& Olsen (2003), and the evolutionary tracks from Maeder \& Meynet (2001)  for  a
metallicity appropriate to the SMC ($z=0.004$).  Solid curves
denote tracks for stars with no initial rotation, while dashed curves show an initial
rotation of 300 km s$^{-1}$.}
\end{figure}

However, one problem had become quite apparent as a result of this work: the location of RSGs in the H-R diagram did {\it not} agree with where the evolutionary tracks
({\it any} evolutionary tracks) said they should be.  Massey \& Olsen (2003) compared
the Magellanic Cloud RSG locations to the Geneva models computed with and
without  ``enhanced" mass loss, as well as to the Padova tracks; Massey (2003a) made
a similar comparison using Galactic RSGs and appropriate metallicity tracks.  In all
cases it appeared that RSGs were significantly cooler and more luminous than the 
tracks would have it.  We show  this for Galactic and SMC RSGs in Figure~\ref{fig:HROLD}.

Now, usually when ``observation" and theory disagree, the problem is with theory,
and indeed when I showed a version of Figure~\ref{fig:HROLD} at the Lanzorte meeting
on massive stars (Massey 2003b), 
Daniel Schaerer was kind enough to point out to me that the
tracks could go further to the right, or not,  depending upon how convection is treated;
this is illustrated in Figure 9 of Maeder \& Meynet (1987).  Still, it seemed as if maybe this
was a case that the ``observations" could be wrong.  We don't ``observe" effective
temperature and bolometric luminosity; instead, we obtain spectra and photometry
and convert these to physical quantities.  Indeed, it turned out that most of the
the basis for the effective temperature scale of RSGs was tied into lunar occultations
of red {\it giants}---there just weren't enough RSGs located along the ecliptic. (See
discussion in Massey \& Olsen 2003.)  {\it If} the effective temperature scales were
actually warmer than what we (and others) had been assuming then that would also
lower the luminosities, yielding better agreement with the tracks, as the bolometric
corrections are quite significant at these cool temperatures.  Of course, if the
effective temperatures turned out to be even cooler than what we had been assuming,
we'd be in even more trouble!

The key to this was using the new generation of MARCS stellar atmospheres (Gustafsson et al.\ 1975, 2003; Plez, Brett, \& Nordlund\ 1992, Plez 2003). An
early version (described by Bessell, Castelli, \& Plez 1998) of these had been used by Oestricher \& Schmidt-Kaler (1998, 1999) to
fit the spectral energy distribution of RSGs in the Magellanic Clouds with some success.
Since that time the models had been improved to include sphericity, with an order of
magnitude increase in the number of opacity sampling points, and revised to incorporate
improved atomic and molecular opacities.

\section{Galactic RSGs}

We first tackled this using a sample of 74 Galactic RSGs. The sample was selected in order to cover the full range of spectral subtypes from early K to the latest M supergiants.
The sample was originally restricted to stars for which there was probable membership in 
OB associations and clusters with known distances (Humphreys 1978; Garmany \& Stencel 1992), although we wound up including a few spectral standards from the
list of Morgan \& Keenan (1973) in order to help with the classifications.  A full
description is given by Levesque et al.\ (2005, hereafter Paper I).  The data covered
the 4000-9000 \AA\ region, with a resolution of 4-6 \AA, and were obtained with the
Kitt Peak 2.1-m telescope and GoldCam spectrometer, and the Cerro Tololo Inter-American Observatory 1.5m telescope and RC spectrometer.

The fitting procedure was straightforward.  We used a grid of MARCS models computed with solar metallicity at 100 K intervals for $\log g$=-1.0 to +1 at 0.5 dex increments.
We interpolated the models to 25 K intermediate values.  A value of $\log g$ of 0.0 is the most physically likely (given the masses and radii of these stars), and we started
with those, letting the depth of the TiO bands then determine the temperature.  We
reddened the models using a Cardelli, Clayton, \& Mathis (1989) extinction law.  Once
we got a good fit by just varying these two parameters (effective temperature and
$A_V$) we then computed the star's location on the HRD, and asked whether or not 
a different $\log g$ would have been more appropriate.  If so, we then repeated the
process.  An uncertainty of 0.5~dex in $\log g$ corresponded to an uncertainty in
$A_V$ of about 0.15~mag, and basically no difference in effective temperature.  We
felt that the latter were determined to a precision of about 50~K or better, except for the earliest
K stars, which lacked TiO bands---for those, the uncertainties were considerably
larger, perhaps as much as 100~K. 

\begin{figure}[!ht]
\plotfiddle{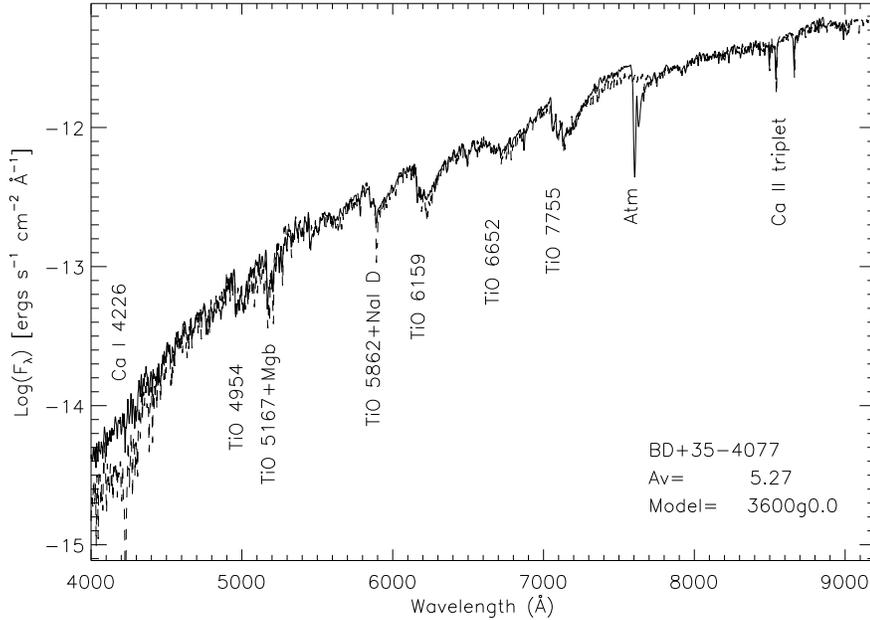}{3.0in}{0.}{70}{70}{-230}{-270}
\caption{\label{fig:fits}An example of our modeling fits.  We show  the spectrum
of BD+35$^\circ$4077, an M2.5 I star, fit with a 3600 K $\log g=0.0$ model reddened
by an $A_V=5.27$ mag.  The observed spectrophotometry is shown as a solid
line, while the model is shown as a dashed line.}
\end{figure}

I want to note, for the record, that the fitting was all done by Emily [Levesque] using
IDL code that she wrote herself, back when she was an undergraduate.  Like all
summer student projects, one never knows whether the whole thing is going to work
out or not.  Would the models be up to the task?  Was the methodology sound?  My
reaction when Emily showed me the first fits was that this was FANTASTIC.  The 
reddened MARCS models not only fit the molecular bands but also fit the continua
far better (I felt) than any models had a right to do with these very difficult stars.  We
show some examples in Figure~\ref{fig:fits}.  Not everything is perfect, of course: for instance, some of
the atomic lines appear to be too strong in the models, and we discuss possible
reasons for this both in Paper I and Levesque et al.\ (2006, hereafter Paper II).

\begin{figure}[!ht]
\plotfiddle{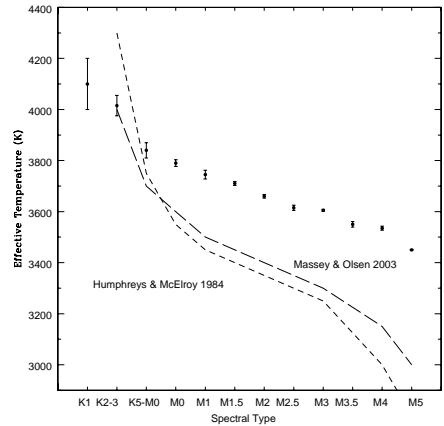}{2.0in}{0.}{30}{30}{-100}{-50}
\caption{\label{fig:teffscale} The new effective temperature scale for Galactic RSGs
is shown by the points and error bars.  The old effective temperature scales of
Humphreys \& McElroy (1984) and Massey \& Olsen (2003) are shown for comparison.}
\end{figure}

When we were done we had new physical properties (effective temperature,
bolometric luminosity, surface gravities, radii) for a large sample of Galactic RSGs.
We can see in Figure~\ref{fig:teffscale} that indeed our work resulted in the effective
temperature scale getting considerably warmer, by about 250 K at spectral type M2.

\begin{figure}[!ht]
\plottwo{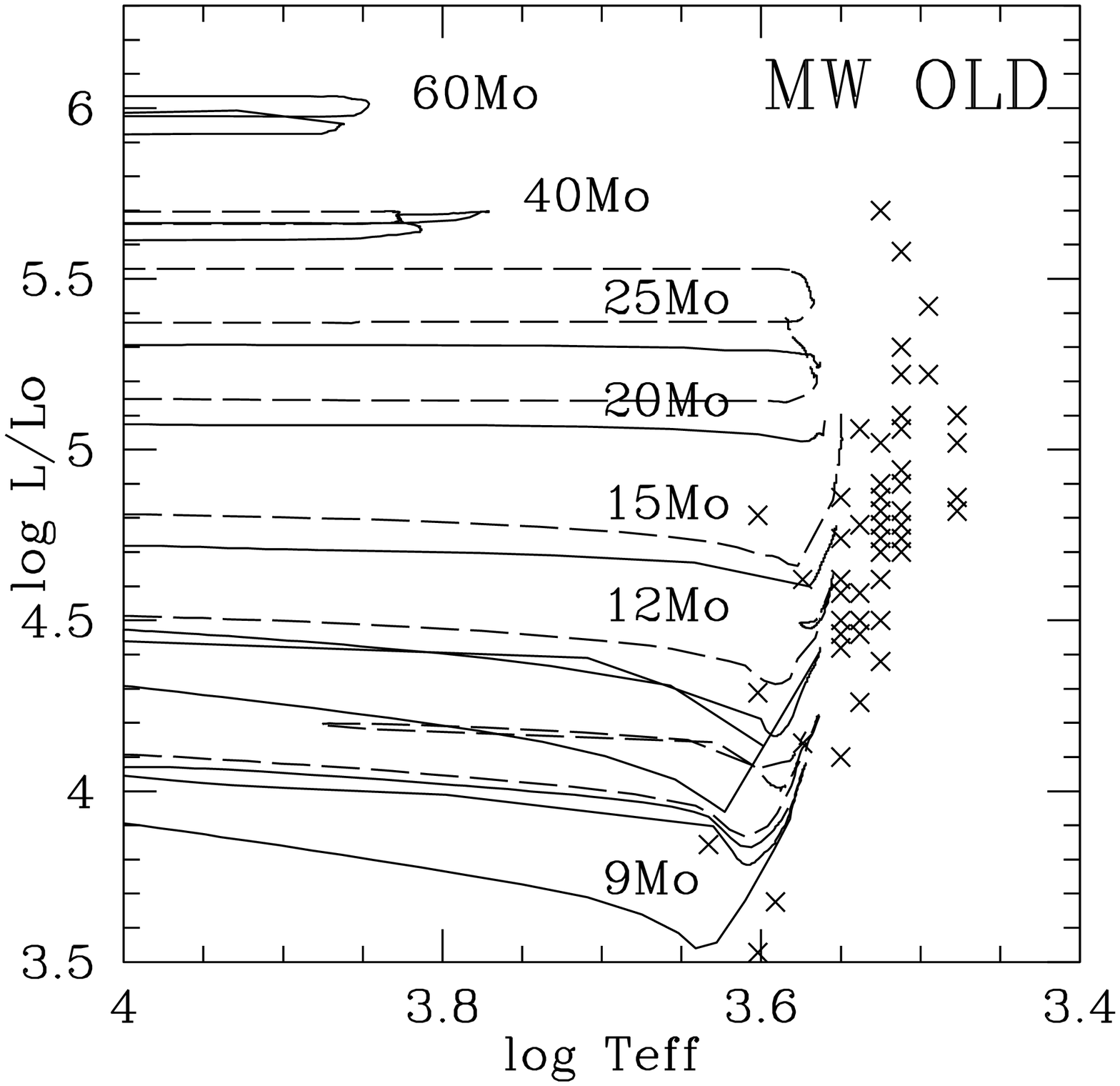}{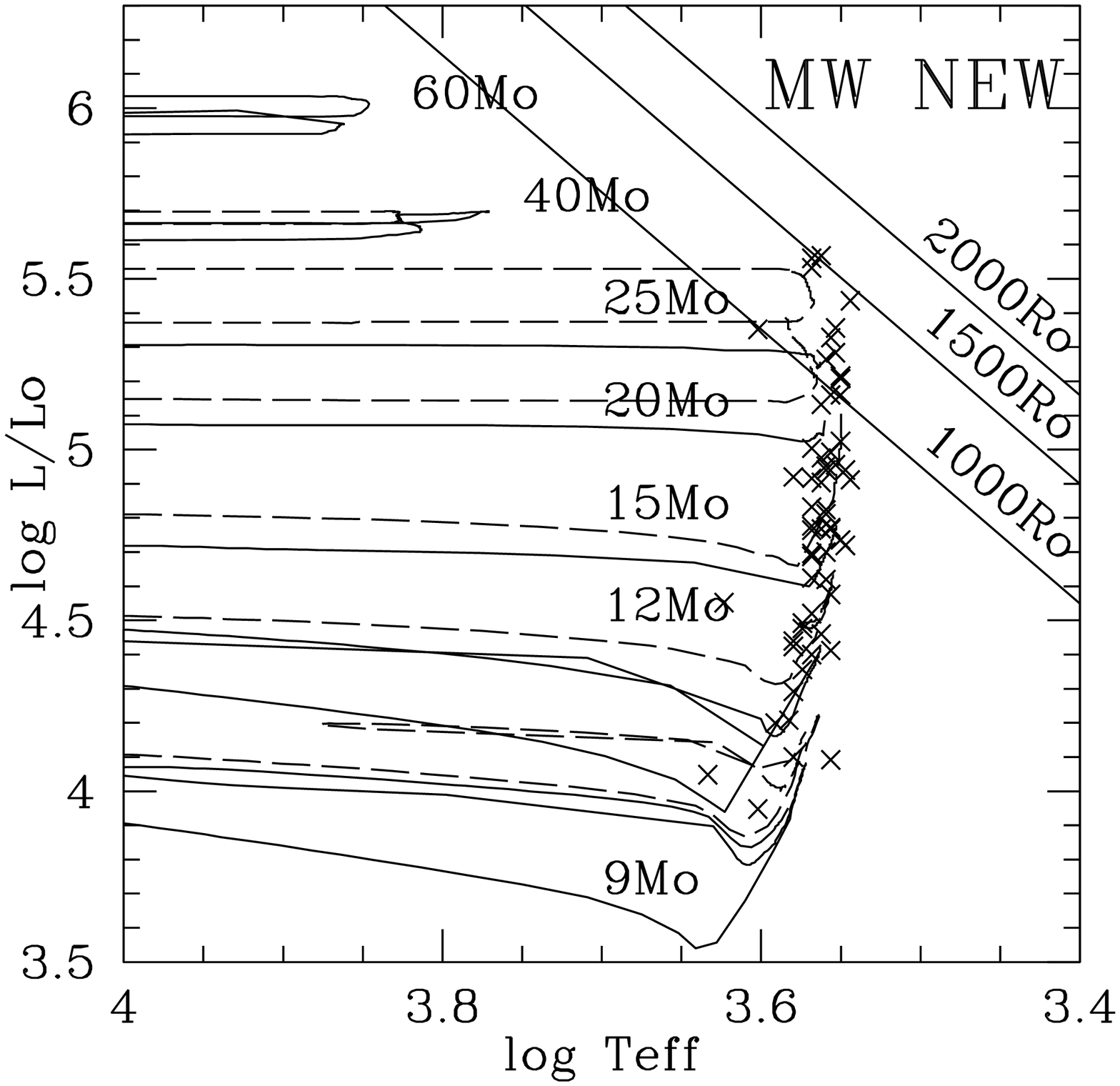}
\caption{\label{fig:HRNEW}  The new effective temperatures and bolometric luminosities
of Galactic RSGs are now in excellent agreement with the models.  On the left
we show the distribution of RSGs in the H-R diagram based upon the old
calibration; on the right, the new.  The
diagonal lines along the upper right denote the radii.}
\end{figure}

What did this do the placement of stars in the H-R diagram?  Exactly what we had
hoped!  We compare the old and new placements in
Figure~\ref{fig:HRNEW}.

We can use this result to answer one other interesting question: {\it How large do
(normal) stars get?}   First, we should ask what do we mean by the "radius" of such
diffuse objects?  Fortunately there's a fundamental definition at hand, namely 
$L=4\pi \sigma T_{\rm eff}^4 R^2$.  Given that, then the radii of the largest stars
are about 1500$R_\odot$, or nearly 7.2 AU!  We found three stars with radii near
that, namely KW Sgr, V354 Cep, and KY Cyg.  The star $\mu$ Cep (Herschel's
``Garnet Star")  comes in a close fourth, at about 1420$R_\odot$.

\section{Extension to Lower Metallicities: Fitting Magellanic Cloud RSGs}

\begin{figure}[!ht]
\plotfiddle{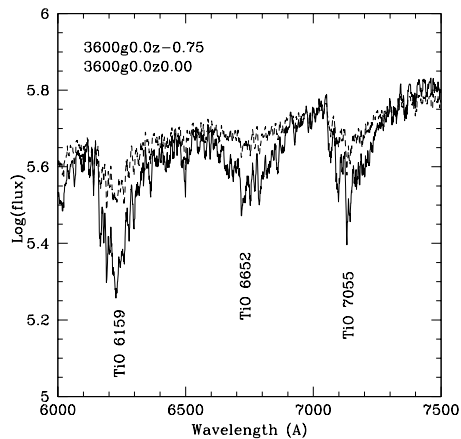}{2.0in}{0.}{30}{30}{-100}{-50}
\caption{\label{fig:wing} A small section of the MARCS models show the differences
expected in TiO band strength with metallicity.  The upper (dotted) model is for a 3600 K
$\log g=0.0$ model with a $\log Z/Z_\odot = -0.75$, similar to that of stars in the SMC.
The lower model has the same effective temperature and surface gravity, but solar
metallicity.}
\end{figure}

Having determined an effective temperature scale for the Milky Way RSGs, and
found excellent agreement with the evolutionary tracks, we were curious to see
what would happen at low metallicities.  After all, the spectral types are determined
by the absolute band strengths of TiO.  So, if there is less TiO available then a star
will have to be cooler to have the same band strength.  At least, this is what we
naively argued, and that appears to be well born out by the models, as shown in 
Figure~\ref{fig:wing}.

We followed the same general procedure as we had in Paper I, first choosing a sample of
RSGs from Massey \& Olsen (2003).   For the Clouds we are faced with a similar problem
recognizing RSGs as we are in M33; Massey \& Olsen had used radial velocities to separate
foreground disk dwarfs from Magellanic Cloud RSGs.  In the end, our study included
36 RSGs in the LMC and 37 in the SMC (Paper II), all observed
on the 4-m Blanco telescope at Cerro Tololo.

The effective temperature scale that we found was indeed cooler than what we found for the Milky
Way for the M-type RSGs, by about 50 K for the LMC and 150 K for the SMC.  The scales for
the K-type RSGs were essentially the same, although our precision in determining temperatures
is certainly worse for early K's (which lack TiO bands) than for the late K's and M's.

\begin{figure}
\plottwo{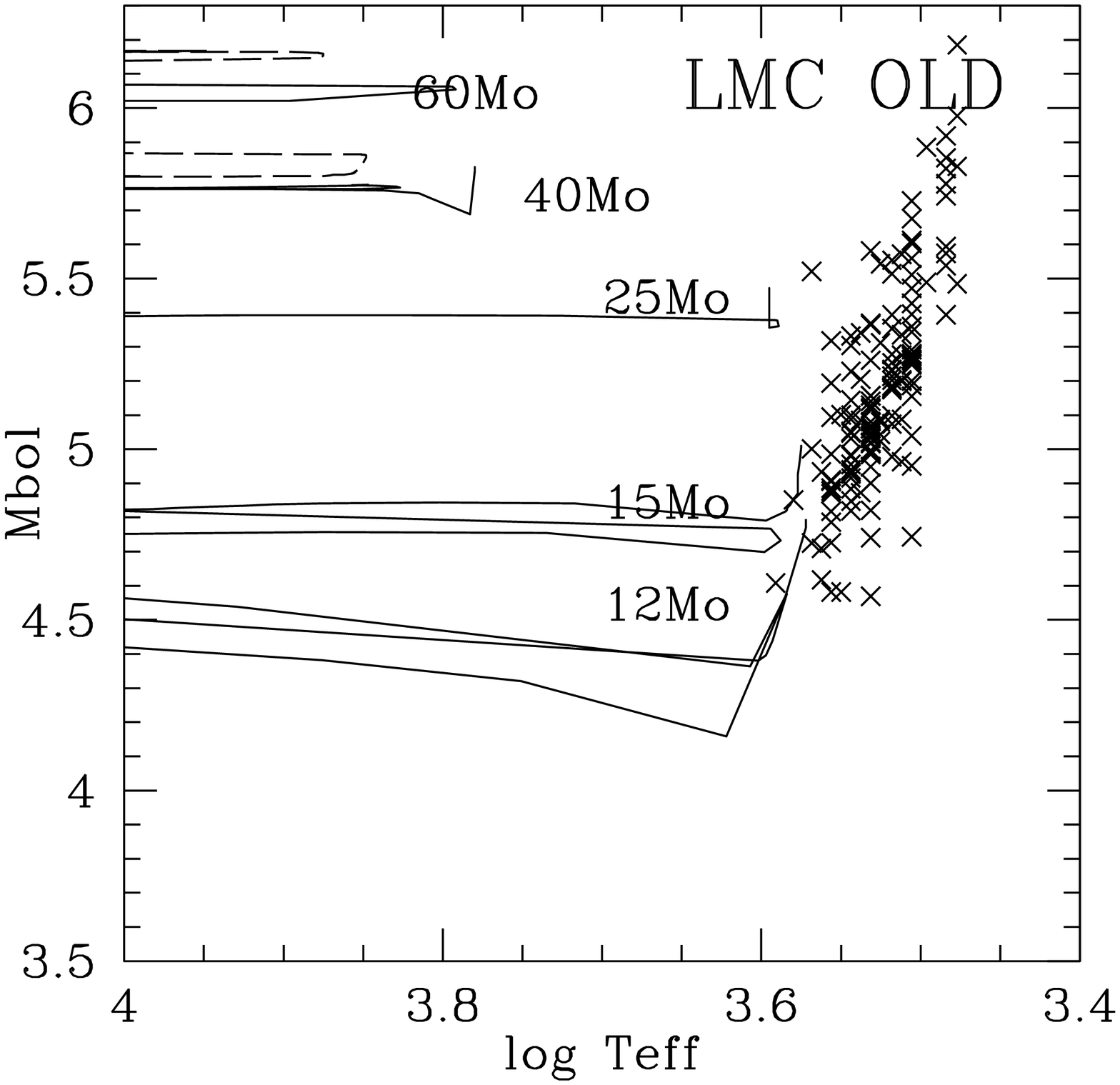}{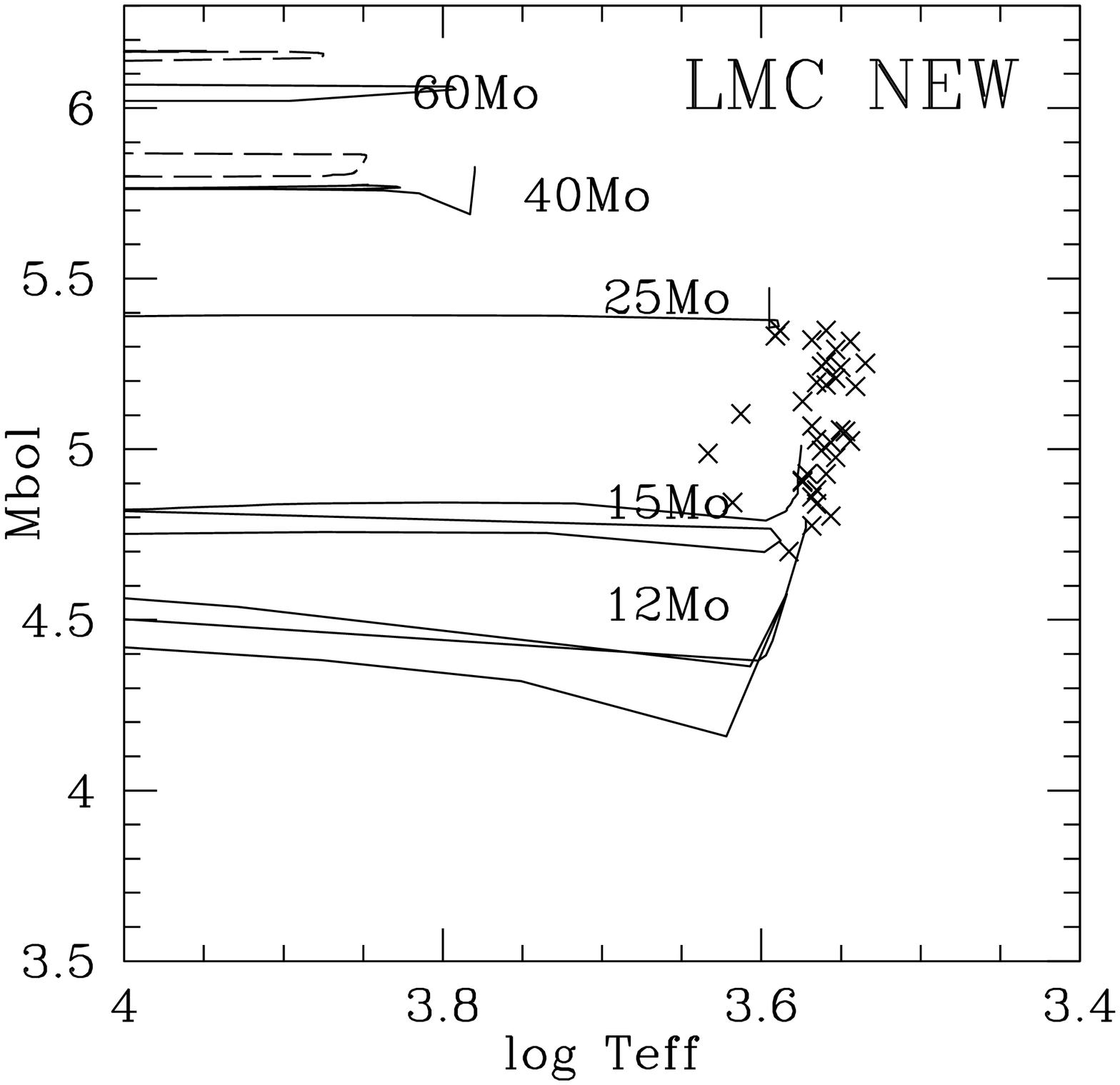}
\caption{\label{fig:LMC}  The H-R diagrams for RSGs in the LMC.  On the left we show the
very poor agreement between the old effective temperature calibration and the evolutionary
tracks; on the right is the MARCS calibration of the effective temperature scale.
Solid lines denote the evolutionary tracks computed for stars with no initial rotation, while
dashed lines are for stars with initial rotation velocities of 300 km s$^{1}$.  The tracks
come from  Meynet \& Maeder (2005). }
\end{figure}

\begin{figure}
\plottwo{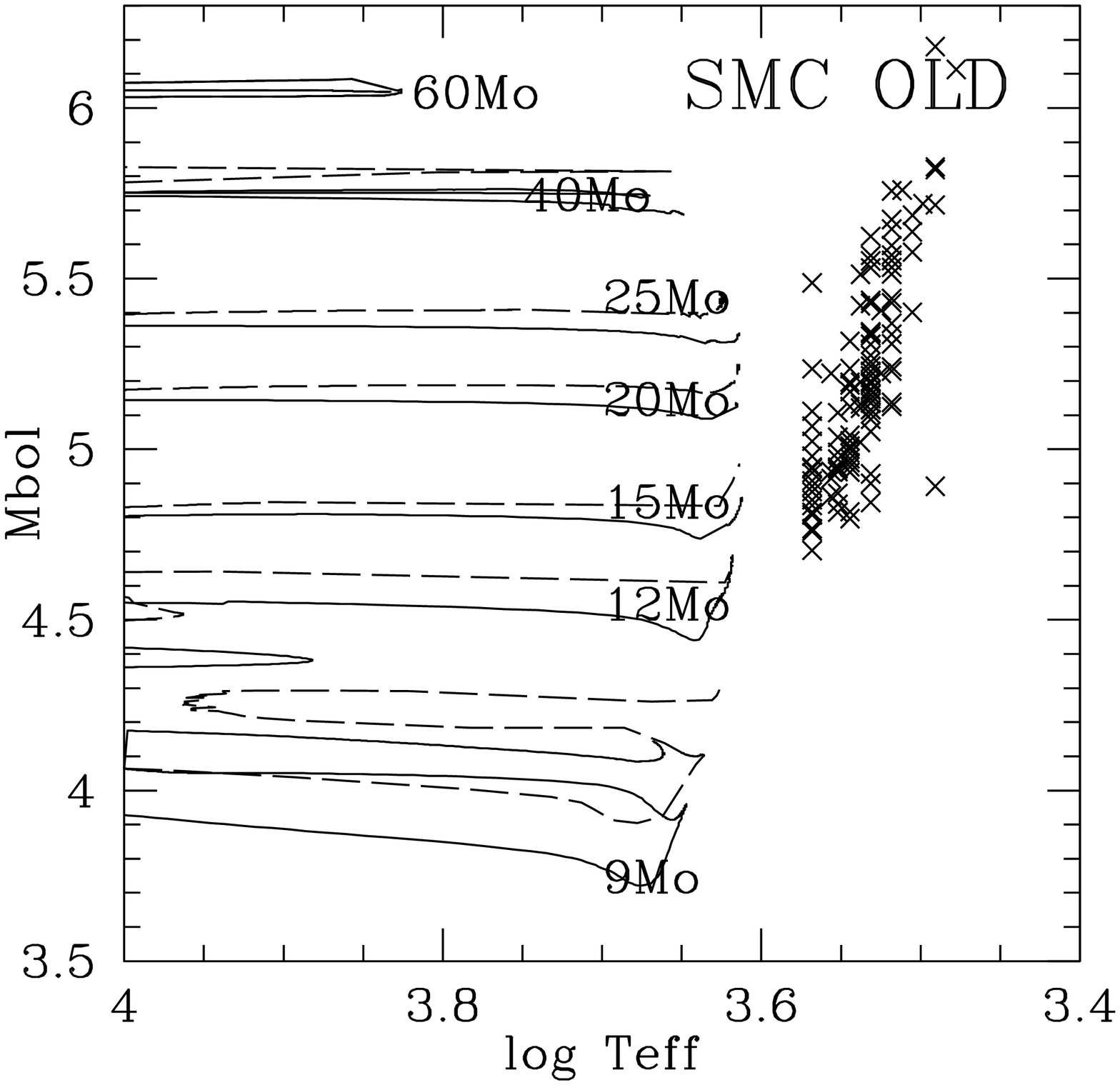}{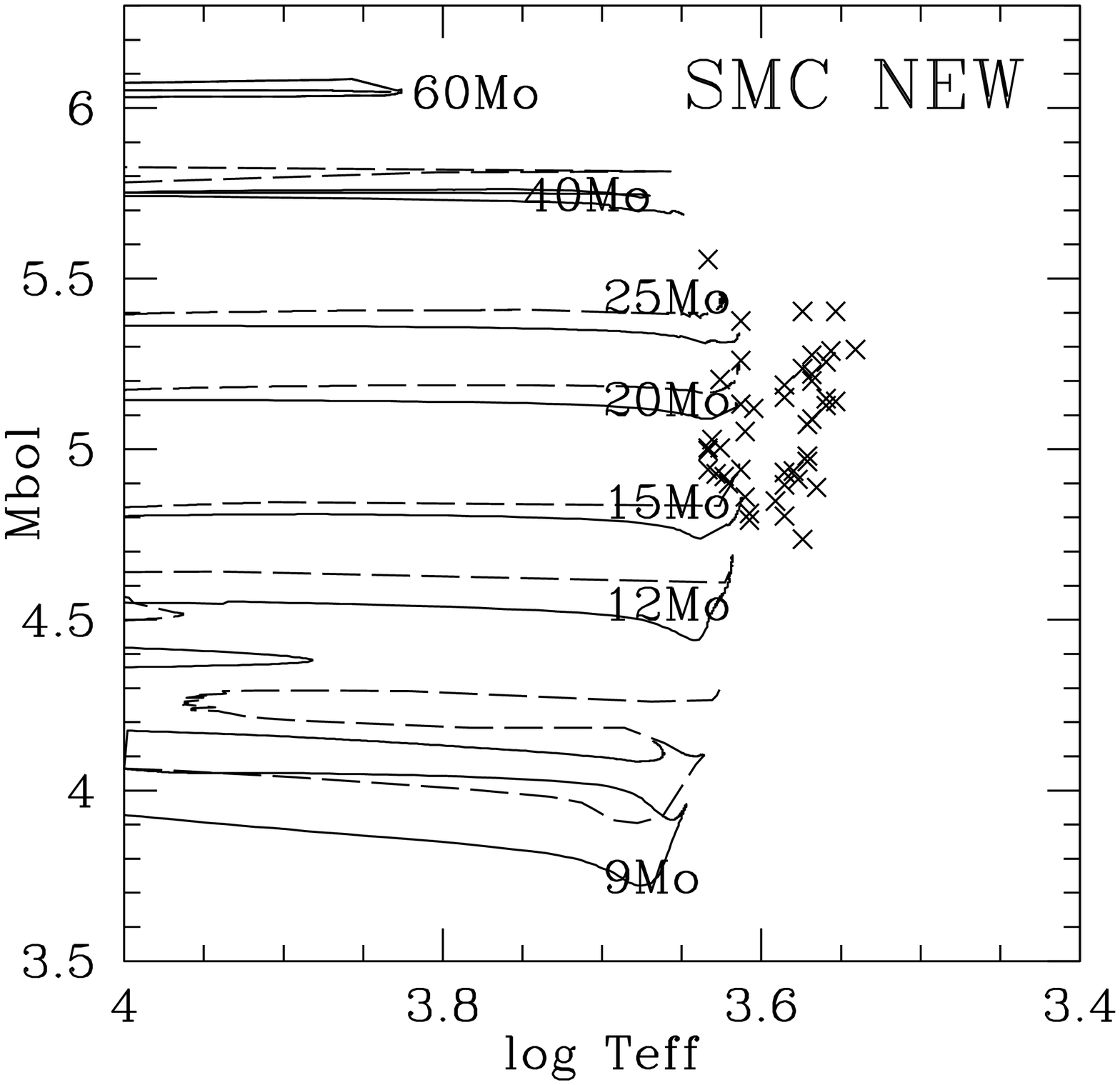}
\caption{\label{fig:SMC}  The H-R diagrams for RSGs in the SMC.
On the left we show the
very poor agreement between the old effective temperature calibration and the evolutionary
tracks; on the right is the MARCS calibration of the effective temperature scale.
Solid lines denote the evolutionary tracks computed for stars with no initial rotation, while
dashed lines are for stars with initial rotation velocities of 300 km s$^{1}$.  The  tracks
come from  Maeder \& Meynet (2001).}
\end{figure}

How did the resulting placement of stars agree with that of the evolutionary tracks?  We show this
in Figures~\ref{fig:LMC} and \ref{fig:SMC}.  
For the LMC the agreement is quite good.  For the SMC there is a
clear improvement, but there is clearly a larger spread in the effective temperatures of RSGs in
the SMC, and some evidence the tracks do not go quite cool enough.  This might be due to
increased importance of rotational mixing in lower metallicity stars (e.g., Maeder \& Meynet 2001).

\section{How Self-Consistent Are the Models?}

We have certainly gotten some nice results with the MARCS models: by fitting the molecular bands
we obtain both effective temperatures and luminosities, with now good agreement with the
evolutionary tracks.  But, we can perform other self-consistency tests. 

\begin{figure}[!ht]
\plottwo{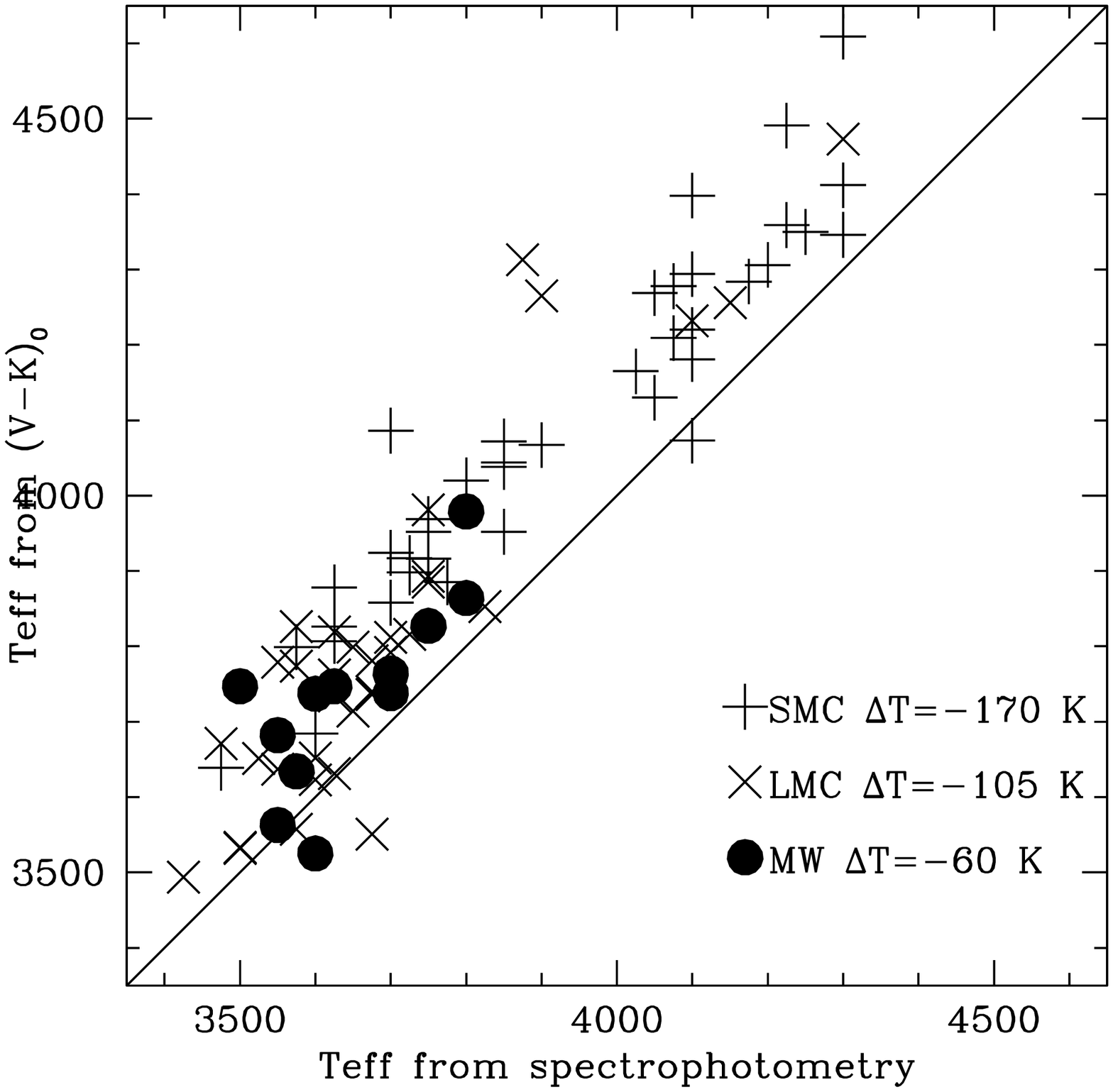}{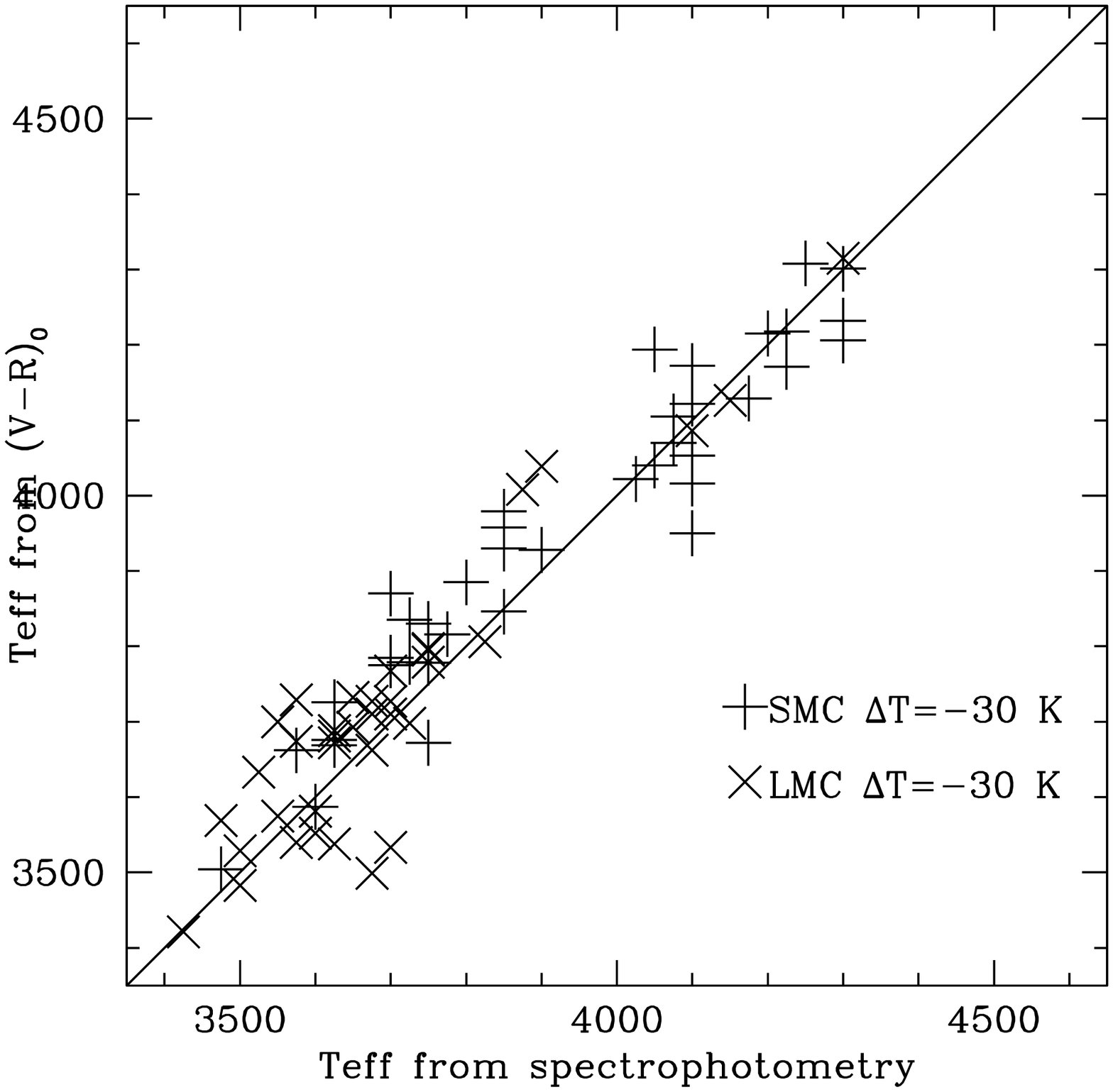}
\caption{\label{fig:broadband} The effective temperatures derived from broad-band colors
compared to those derived from spectral fitting.  On the left we show the comparison for
$(V-K)_0$; on the right, for $(V-R)_0$.  Solid dots correspond to the Milky Way, ``+" symbols
for the LMC, and ``x" symbols for the SMC.}
\end{figure}

One test is to separately derive the physical properties of these stars using the broad-band colors
of the stars.  We chose for this test the $(V-K)_o$ and $(V-R)_o$ colors.  Now, this is not completely
independent, as we have to assume the color excesses $E(B-V)$ from our fitting of the spectrophotometry.   Still, it provides an important
 check on whether the spectral flux distribution of the models
are consistent with the TiO band strengths of the models.  We show the comparison for $(V-K)_0$
in Figure~\ref{fig:broadband}.  
Solid points are for the Milky Way, ``+" signs are for the LMC, and ``x" are for
the SMC.  And, we do find some systematic differences.  We find warmer effective temperatures (and
hence smaller luminosities) using just the $(V-K)_0$ colors.  The differences appear to be metallicity
dependent, with an median $\Delta T=-60$~K for the Milky Way, -105~K for the LMC, and -170~K for the
SMC.  If we make the same comparison, though, for $(V-R)_0$ the agreement is excellent.

The explanation we offered (in Paper II) is that this may just be a reflection of the intrinsic 
limitations of static, 1-D models.  Ryde et al.\ (2006) found that MARCS models at the 
canonical 3600 K effective temperature for Betelgeuse do not reproduce the IR H$_2$O
lines---instead, a much lower temperature, 3250 K, is needed.
Radiative-hydrodynamic 3-D models (Freytag et al.\ 2002) show that there are likely
warm and cool patches on the surfaces of these stars, which may explain such wavelength
dependent
effects.

\section{When Smoke Gets in Your Eyes}

One of the truly surprising things we found---although really we should have expected it---is that
many of these RSGs have a significant amount of circumstellar reddening, presumably due to
dust.  We give the arguments in Massey et al.\ (2005), and let me repeat them here.  

\begin{figure}[!ht]
\plotfiddle{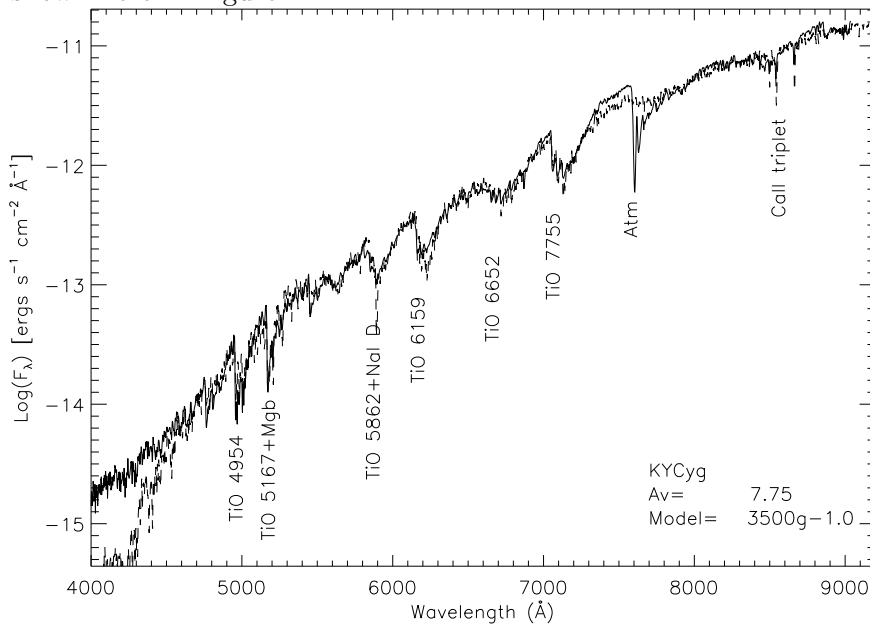}{3.0in}{0.}{70}{70}{-230}{-250}
\caption{\label{fig:nuv}An example of the near-UV problem.  We show  the spectrum
of KY Cyg, an M3-4 I star, fit with a 3500 K $\log g=-1.0$ model reddened
by an $A_V=7.75$ mag.  The observed spectrophotometry is shown as a solid
line, while the model is shown as a dashed line.}
\end{figure}

Our first clue that there was something funny going on was in our Galactic study: although the 
spectral energy distributions were generally well fit by the models, there were a number of stars
that showed significant discrepancies in the wavelength region $<4000$\AA, always in the sense
that the stars showed more flux than the models predicted. Look carefully at Figure~\ref{fig:fits}.
A more drastic case is shown here in Figure~\ref{fig:nuv}.

We considered
a number of possibilities.  Were these stars binaries?  Well, we had found, and eliminated, a few
stars from our original sample because Balmer lines were present.  We convinced ourselves that
the binary explanation did not work for the other stars as there were no signs that the spectra
were composite. We worried a lot about instrumental effects; after all, $F_{\lambda 7000}/F_{\lambda 3500}$
is about 10,000 for our most heavily reddened RSGs, while it is roughly unity for our spectrophotometric
standards.  We eliminated this explanation by  sticking in a CuSO$_4$ blocking filter (which 
should not be needed at all in first order) and obtaining the same results. We could imagine 
theoretical explanations for this discrepancy: as discussed above, hot spots may be present on the
surfaces of these stars (Freytag et al.\ 2002).  Additionally, chromospheric emission might play a role
(Carpenter et al.\ 1994; Harper et al.\  2001).  

However, we also noticed something odd: the stars with the greatest near-UV discrepancy also seemed
to be the stars with the highest reddening.  We looked at this most closely.  In general, these RSGs
were considerably {\it more} reddened than the OB stars in the {\it same} clusters and associations.
What was the deal with that?  The obvious explanation seemed to be circumstellar dust.  We know that
RSGs are ``smoky" in the sense that dust condenses out of the stellar winds at radii of 5-10$R_{\rm star}$, i.e, about 1000$R_\odot$.    The presence of circumstellar dust shells was first revealed by
ground-based IR photometry (Hyland et al.\ 1969), while {\it IRAS} two-color diagrams established that
such shells were a nearly ubiquitous phenomenon (Stencel et al.\ 1988, 1989).   Such dust production
is thought to be episodic, with timescales of a few decades (Danchi et al.\ 1994)

In retrospect, it is surprising that someone had not thought of this earlier.   In Massey et al.\ (2005)
we did a simple thin-shell approximation and found that in fact one would {\it expect} RSGs to have
of order 1~mag of visual extinction for average dust production rates over a 10-yr time span.  Stars
with stronger winds (more luminous) or whose dust condenses closer to the star should have more, while ones with weaker winds (less luminous) or whose dust condenses further from the star will have less.  Such dust is likely to have a different grain size distribution that is standard in the ISM and
thus might scatter differently in the NUV, but regardless of this we would expected to see more NUV
light simply because blue light would be scattered into the beam by
parts of the unresolved circumstellar dust shell off-axis to the line of sight.

Our work established that there was a pretty good correlation of dust production rates with luminosity,
and so we were able to obtain an estimate for the contribution of dust from RSGs to the ISM. In
the solar neighborhood this value is about $3\times 10^{-8} M_\odot$ yr$^{-1}$ kpc$^{-2}$.  This is
about 1\% of the return rate of AGBs (Whittet 2003).  However, in galaxies at large look-back times (where AGBs have not yet had time to form in large numbers), or in metal-poor starbursts, we expect
that the dust produced by RSGs will dominate.

\section{What's Next?}

The obvious extension of our work is to extend it now to lower and high metallicities.  Fortunately,
that's relatively easy (and fun!) to do:

\subsection{M31: Where the Metallicity Really is Higher}

 M31 (the Andromeda galaxy) has long been supposed to have a metallicity roughly twice solar,
 based upon nebular analysis of its H~II regions (see, for example, Zaritzky et al.\ 1994).  Here
 we would expect to find that RSGs were somewhat colder (at a given spectral type) than their
 Galactic or Magellanic Cloud counterparts, and that the average spectral types are even later.
 We used the Local Group Galaxy Survey data (Massey et al.\ 2006) to select a sample of
 RSG candidates (using a $B-V$ vs.\ $V-R$ two-color diagram) and then used the WIYN 3.5-m
 and the Hydra fiber positioner to determine radial velocities to establish who was really an M31
 RSG.  The MMT 6.5-m was then used for spectroscopy.  
 
\begin{figure}[!ht]
\plottwo{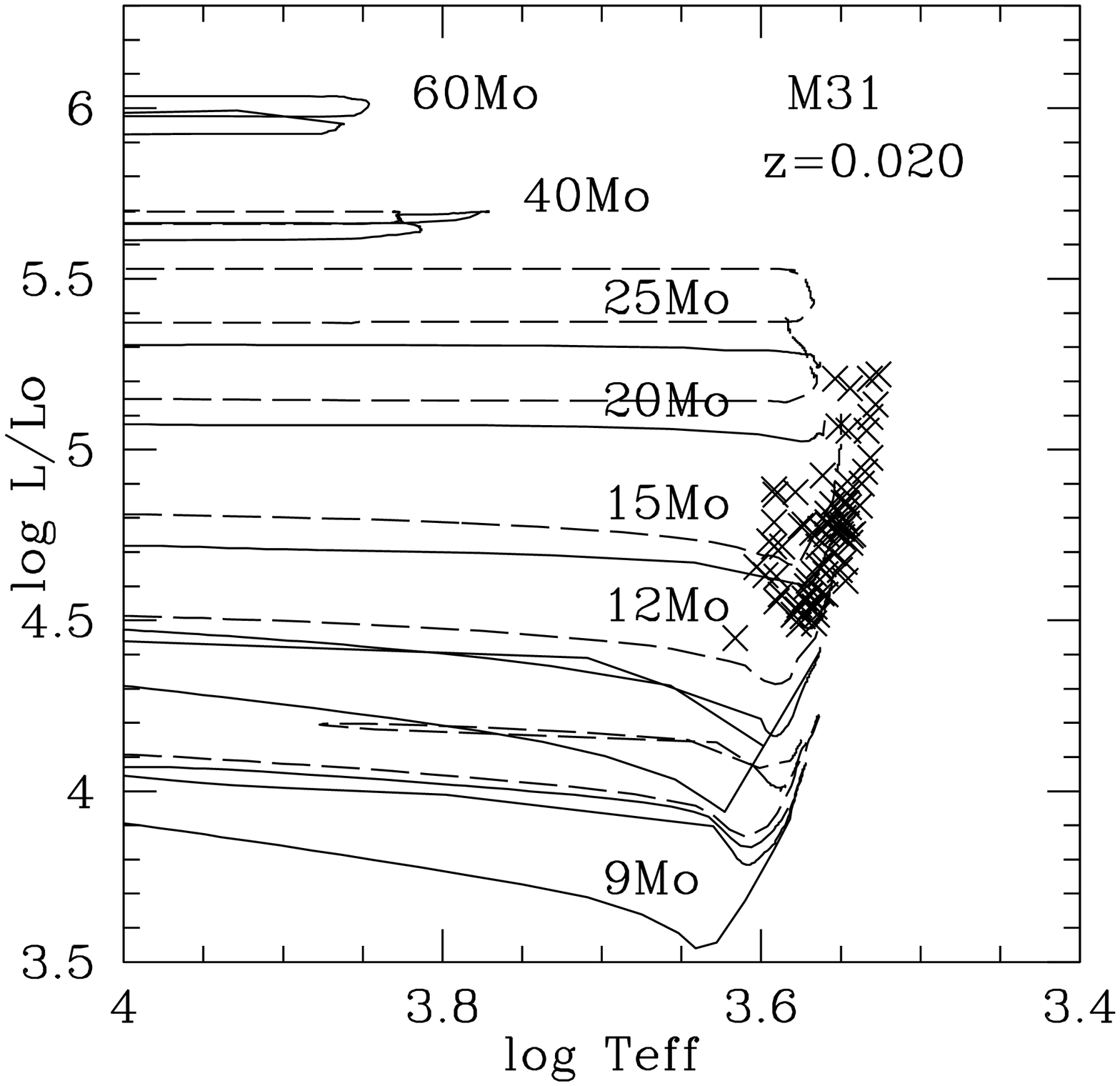}{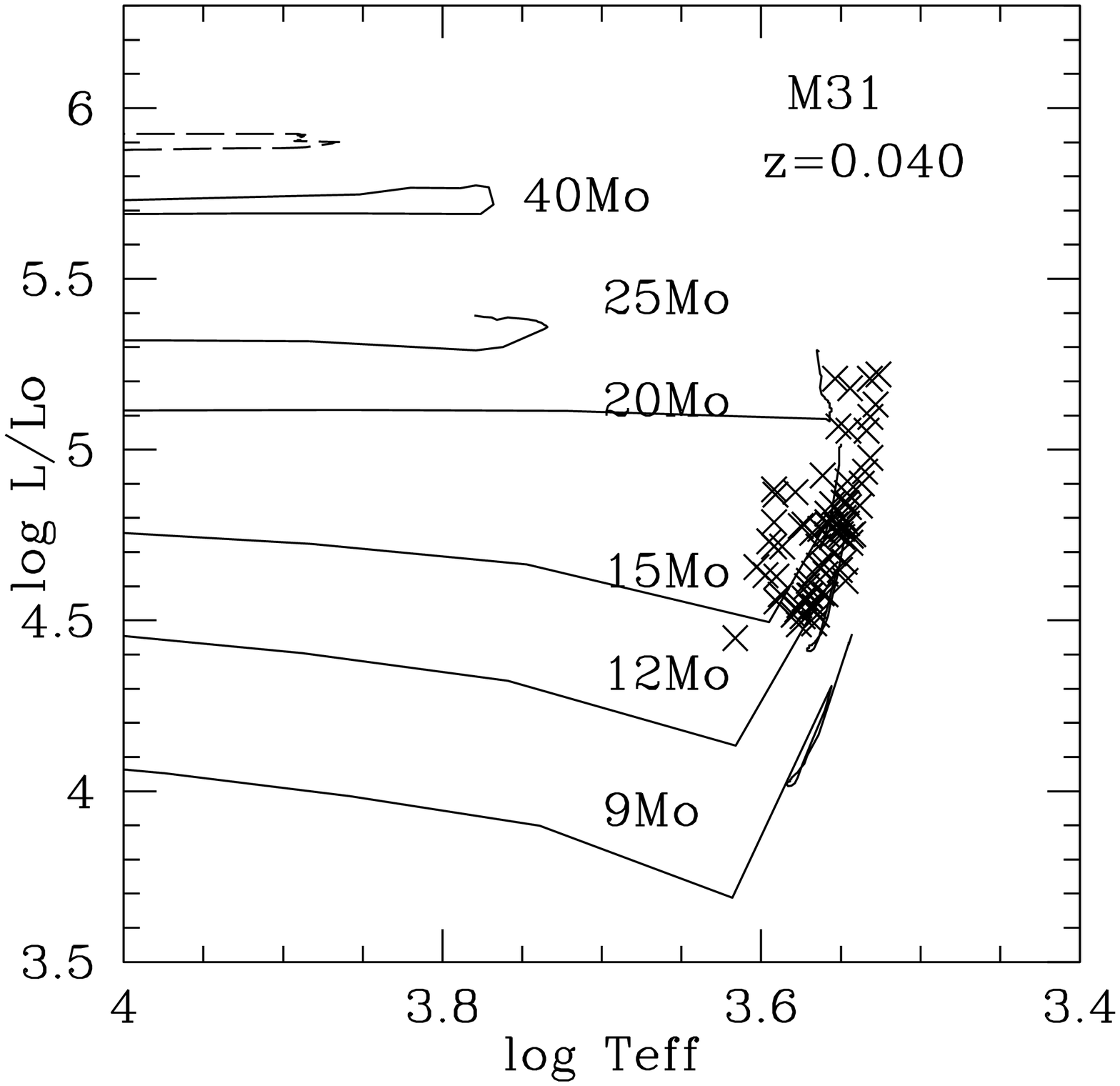}
\caption{\label{fig:M31hrds}. RSGs in M31.  We show here two H-R diagrams based upon 
2MASS and LGGS $V-K$.  On the left we have used the MARCS atmospheres to determine 
effective temperatures based upon a solar metallicity calibration and compare these to the
solar-metallicity ($z=0.020$) metallicity evolutionary tracks from  Meynet \& Maeder (2003).
On the right we have repeated the experiment for a metallicity of 2 times solar, corresponding to
the metallicity determined from HII regions.  If the metallicity were really solar, then there should
be a lot of higher luminosity RSGs than what we actually find: in the diagram on the left, we see
we would expect to see RSGs with luminosities of $\log L/L_\odot =5.5$, and we don't.  Instead, the
highest luminosities we see are $\log L/L_\odot=5.2$, which is all we expect from the higher
metallicity tracks shown on the right.}
\end{figure}

 We are still in the process of analysis, but one thing is already certain: M31 really does have 
 (something like) two times solar metallicity.  This fact had been called into question by Smartt
 et al.\ (2001), who did a chemical analysis of a single B-type supergiant in that galaxy, finding
 nearly solar-type abundances.
 In Figure~\ref{fig:M31hrds} we show that if the abundances really were {\it solar} then there
 should be a lot more luminous RSGs than what we actually see.
 
 \subsection{WLM : Where the Metallicity is Extremely Low}
 As this text is being written, I am in Chile preparing to observe RSGs in WLM, a galaxy with the
 lowest metallicity of any galaxy in the Local Group that is forming massive stars.  Will we find
 even worse agreement between the evolutionary tracks and the location of RSGs in the 
 H-R diagram?  Will there be peculiarly late RSGs that change their spectral types from year
 to year, as we found in the SMC (Massey et al.\ 2007a, Levesque et al.\ 2007, plus Levesque
 et al.\ in this volume).  As our colleague Nidia Morrell said when she saw the spectrum of HV 11423,
 ``That star is in a lot of trouble!"
 
 \acknowledgements 

 Being an astronomer working on RSGs is very fun and exciting---we don't know what we're going to
 find next---and I've appreciated the opportunity to come to this well-organized meeting and learn a lot
 from cool star experts.    Part of the work described today was done in collaboration with Georges Meynet, Andre Maeder, and Eric Josselin.  This work is support through the National Science Foundation AST-0604569.




\begin{references}
\reference {} Bessell, M. S., Castelli, F., \& Plez, B. 1998, A\&A, 333, 231
\reference {} Carpenter, K. G., Robinson, R. D., Wahlgren, G. M., Linsky, J. L., \& Brown, A. 1994, ApJ, 428, 329
\reference {} Cardelli, J. A., Clayton, G. C., \& Mathis, J. S. 1989, ApJ, 345, 245
\reference {} Danchi, W. C., Bester, M., Degiacomi, C. G., Greenhill, L. J., \& Townes, C. H. 1994, AJ, 107, 1469
\reference {} Freytag, B., Steffen, M., \& Dorch, B. 2002, Astron.\ Nachr., 323, 213
\reference {} Garmany, C. D., \& Stencel, R. E. 1992, A\&AS, 94, 211
\reference {} Gustafsson, B., Bell, R. A., Eriksson, K., \& Nordlund, \AA. 1975, A\&A, 42, 407
\reference {} Gustafsson, B., Edvardsson, B., Eriksson, K., Mizumo-Wiedner, M., Jorgensen, U. G., \& Plez, B. 2003, in ASP Conf.\ Ser.\ 288, Stellar Atmosphere Modeling, ed.\ I. Hubeny, D. Mihalas, \& K. Werner (San Francisco: ASP), 331
\reference {} Harper, G. M., Brown, A., \& Lim, J. 2001, ApJ, 551, 1073
\reference {} Hyland, A. R., Becklin, E. E., Neugebauer, G., \& Wallerstein, G. 1969, ApJ, 158, 619
\reference {} Humphreys, R. M. 1978, ApJS, 38, 309
\reference {} Humphreys, R. M., \& McElroy, D. B. 1984, ApJ, 284, 565
\reference {} Humphreys, R. M., \& Sandage, A. 1980, ApJS, 44, 319
\reference {} Levesque, E. M., Massey, P., Olsen, K. A. G., Plez, B., \& Josselin, E., Maeder, A., \& Meynet, G. 2005, ApJ, 628, 973 (Paper I)
\reference {} Levesque, E. M., Massey, P., Olsen,  K. A. G., Plez, B., Meynet, G., \& Maeder, A. 2006, ApJ, 645, 1102 (Paper II)
\reference {} Levesque, E. M., Massey, P., Olsen, K. A. G., \& Plez, B. 2007, ApJ, 667, in press (astro-ph/0705.3431
\reference {} Maeder, A., Lequeux, J., \& Azzopardi, M. 1080, A\&A, 90, L17
\reference {} Maeder, A., \& Meynet, G. 1987, A\&A, 182, 243
\reference {} Maeder, A., \& Meynet, G. 2001, A\&A, 373, 555
\reference {} Massey, P. 1998, ApJ, 501, 153
\reference {} Massey, P. 2003a, ARA\&A, 41, 15
\reference {} Massey, P. 2003b, in A Massive Star Odyssey: From Main Sequence to Supernova, IAU Symp.\ 212, ed. K. van der Hucht, A. Herrero, \& C. Esteban
(San Francisco: ASP), 316
\reference {} Massey, P., \& Olsen, K. A. G. 2003, AJ, 126, 2867
\reference {} Massey, P., Levesque, E. M., Olsen, K. A. G., Plez, B., \& Skiff, B. A. 2007a, ApJ, 660, 301
\reference {} Massey, P., Olsen, K. A. G., Hodge, P. W., Jacoby, G. H., McNeill, R. T., Smith, R. C., \& Strong, S. B. 2007b, AJ, 133, 2393
\reference {} Massey, P., Olsen, K. A. G., Hodge, P. W., Strong, S. B., Jacoby, G. H., Schlingman, W., \&
Smith, R. C  2006, AJ, 131, 2478
\reference {} Massey, P., Plez, B., Levesque, E. M., Olsen, K. A. G., Clayton, G. C., \& Josselin, E. 2005,
ApJ, 634, 1286
\reference {} Meynet, G., \& Maeder, A. 2003, A\&A, 404, 975
\reference {} Meynet, G., \& Maeder, A. 2005, A\&A, 429, 581
\reference {} Morgan, W. W., \& Keenan, P. C. 1973, ARA\&A, 11, 29
\reference {} Oestreicher, M. O., \& Schmidt-Kaler, T. 1998, MNRAS, 299, 625
\reference {} Oestreicher, M. O., \& Schmidt-Kaler, T. 1999, Astron.\ Nachr.\ 320, 385
\reference {} Plez, B. 2003, in ASP Conf.\ Ser.\ 298, {\it GAIA} Spectroscopy: Science and
Technology,  ed. U. Munari (San Francisco: ASP), 189
\reference {} Plez, B., Bett, J. M., \& Nordlund, \AA\ 1992. A\&A, 256, 551
\reference {} Ryde, N., Harper, G. M., Richter, M. J., Greathouse, T. K., \& Lacy, J. H. 2006, ApJ, 637,
1040
\reference {} Smartt, S. J., Crowther, P. A., Dufton, P. L., Lennon, D. J., Kudritzki, R. P., Herrero, A.,
McCarthy, J. K., \& Bresolin, F. 2001, MNRAS, 325, 257
\reference {} Stencel, R. E., Pesce, J. E., \& Bauer, W. H. 1988, AJ, 95, 141
\reference {} Stencel, R. E., Pesce, J. E., \& Bauer, W. H. 1989, AJ, 97, 1120
\reference {} Zaritzky, D., Kennicutt, R. C., \& Huchra, J. P. 1994, ApJ, 420, 87

\end{references}
\end{document}